\documentclass[aps, prb, twocolumn, amsmath, amssymb, showpacs]{revtex4-1}

\usepackage[utf8]{inputenc}
\usepackage{graphicx}
\usepackage{units}
\usepackage{color}
\usepackage[pdftex, colorlinks=true, linkcolor=myblue, citecolor=myblue]{hyperref}
\usepackage{txfonts}

\definecolor{myblue}{rgb}{0.1,0.,.85}

\begin{document}

\title{Lifetime of the surface magnetoplasmons in metallic nanoparticles}

\author{Guillaume Weick} 
\author{Dietmar Weinmann}
\affiliation{Institut de Physique et Chimie des Mat\'eriaux de Strasbourg (UMR
7504), CNRS and Universit\'e de Strasbourg, 23 rue du Loess, BP 43, F-67034
Strasbourg Cedex 2, France}


\begin{abstract}
We study the influence of an external magnetic field on the collective electronic
excitations in metallic nanoparticles. While the usual surface plasmon
corresponding to the collective oscillation of the electrons with
respect to the ionic background persists in the direction parallel to the
magnetic field, the components in the perpendicular plane are affected by the
field and give rise to two collective modes with
field-dependent frequencies, the surface magnetoplasmons.
We analyze the decay of these collective excitations by their coupling to
particle-hole excitations and determine how their lifetimes are modified  by the
magnetic field. 
In particular, we show that the lifetime of the usual surface plasmon 
is not modified by the magnetic field, while the
lifetime of the two surface magnetoplasmons 
present a weak magnetic-field dependence. 
Optical spectroscopy experiments are suggested in
which signatures of the surface magnetoplasmons may be observed.
\end{abstract}

\pacs{73.20.Mf, 73.22.Lp, 78.67.Bf}

\maketitle

\section{Introduction}
\label{sec:introduction}
Collective excitations in confined many-body systems are of great fundamental
interest. Such excitations decay due to their 
coupling to other internal degrees of freedom of the system. 
This allows the study of quantum dissipation and
decoherence and thereby the transition between quantum and
classical physics. 
Particularly well studied is the case of surface plasmon excitations in metallic 
nanoparticles. \cite{bertsch, dehee93_RMP, brack93_RMP, kreibig}
These collective dipolar vibrations of the electronic center of mass with
respect to the ionic background dominate the 
optical absorption 
spectrum. The corresponding 
resonance linewidth gives indirect access to the lifetime of the surface
plasmon. Moreover, pump-probe experiments allow one to follow the time evolution of
the electron dynamics 
with a resolution of a few femtoseconds, \cite{bigot95_PRL, bigot00_CP, delfa00_CP} 
and the surface plasmon excited by the probe laser field plays a prominent role
in the interpretation of the pump-probe measurements. \cite{weick07_EPL}
Recently, this technique was combined with magneto-optical Kerr effect 
measurements to follow the magnetization dynamics in superparamagnetic 
transition-metal nanoparticles and in particular the ultrafast demagnetization resulting from 
the pump laser excitation. \cite{andra06_PRL} This demagnetization is also observed 
in ferromagnetic thin films, \cite{vomir05_PRL} and its explanation is still a 
matter of debate. \cite{bigot09_NaturePhys, kiril10_RMP}

\begin{figure}[tb]
\includegraphics[width=\linewidth]{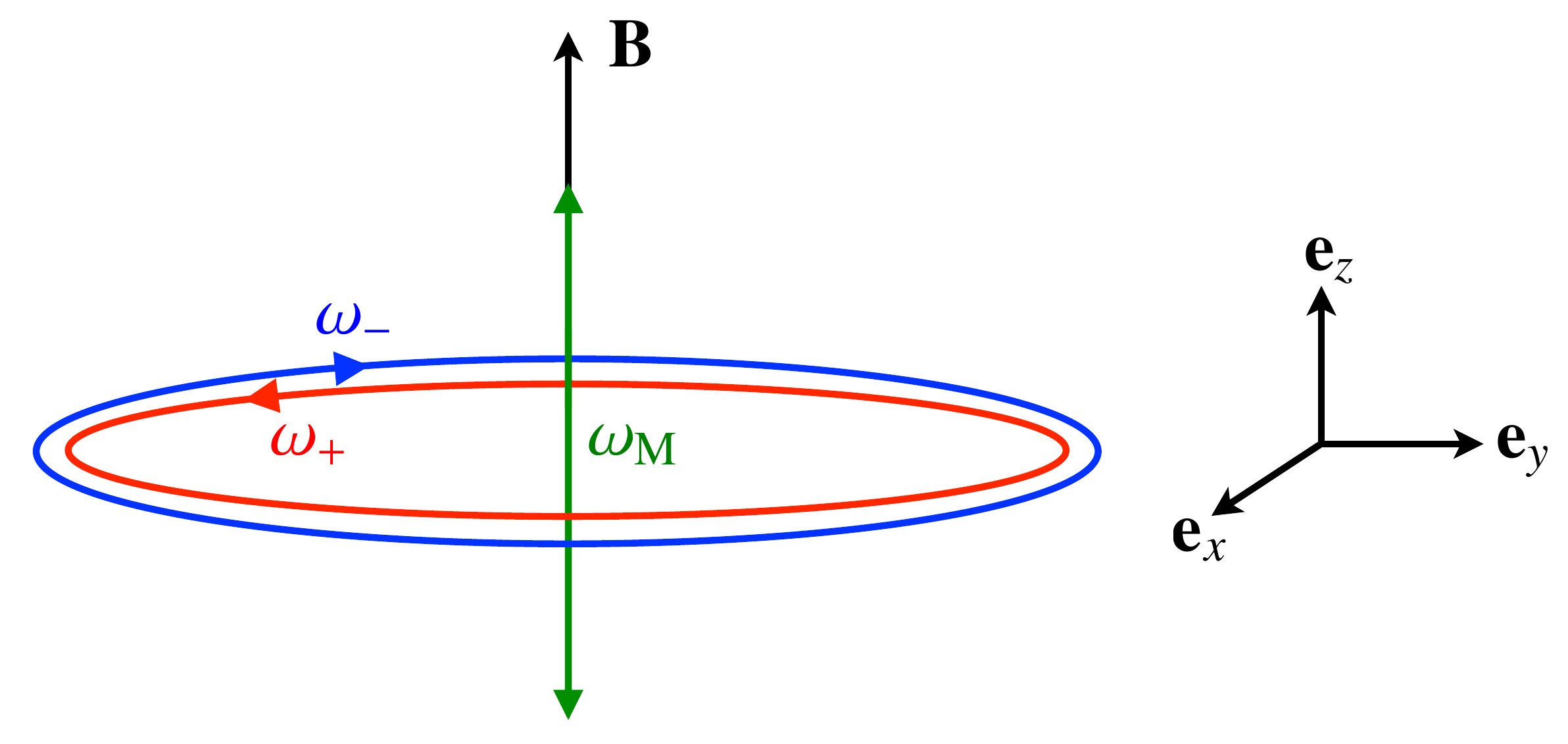}
\caption{\label{fig:sketch}%
(Color online) Sketch of the center-of-mass motion with frequency $\omega_\mathrm{M}$ 
for the surface plasmon collective mode parallel 
to the magnetic field $\mathbf{B}$, and for the two surface magnetoplasmon modes
with frequencies
$\omega	_+$ and $\omega_-$ [cf.\ Eq.~\eqref{eq:omega_pm}], where it rotates
counterclockwise and clockwise in the plane
perpendicular to $\mathbf{B}$, respectively.}
\end{figure}

The saturation magnetization in ferromagnetic materials
plays the role of an effective \textit{external} magnetic field that couples to the 
orbital degrees of freedom. \cite{kitte63_PRL} 
It is therefore relevant to study the influence of this effective
field on the collective resonances in nanoparticles. 
In this work, we focus on the generic problem of the role played by 
an external magnetic field on the
collective resonances in spherical \textit{nonmagnetic} (i.e., alkaline or noble-metal) 
nanoparticles. While the surface 
plasmon excitation with the dipole parallel to the magnetic field is not 
modified, the two plasmon modes with dipoles perpendicular to the field evolve 
in two magnetic-field-dependent collective modes when the magnetic field is 
switched on (see Fig.~\ref{fig:sketch}). 
We term these collective excitations ``surface 
magnetoplasmons." They are split in energy by an amount of the order of the 
cyclotron frequency $\omega_\mathrm{c}$. 
Similar excitations exist in quasi-two-dimensional semiconductor quantum dots. 
\cite{jacak, reima02_RMP} 
In the case of quantum dots, the energy scale of the resonance energies is not 
in the optical but in the infrared range, and 
the coupling of the collective excitation to the other 
degrees of freedom is very weak. \cite{sikorski,meurer}
In contrast, in metallic nanoparticles, this coupling is quite strong and 
leads to the decay of the collective modes. 
It is the aim of the present paper to analyze in detail the consequences of such
a coupling on the lifetime of the surface magnetoplasmons.
As we show in the sequel, the coupling yields a non-negligible
linewidth of the
corresponding resonances which might limit their observability in experiments.

In this work, we analyze the decay of the surface magnetoplasmons in metallic nanoparticles
which is caused by  
their coupling to particle-hole excitations. This process, called Landau
damping,
\cite{bertsch} is the dominant decay channel for intermediate-size nanoparticles
with a radius $a$ in the range $\unit[1]{nm}\lesssim a\lesssim\unit[10]{nm}$.
\cite{footnote:radiation}
We show that the surface magnetoplasmon lifetimes (whose inverses yield the
linewidths of the resonances) increase linearly with the 
size of the nanoparticle, as it is the case for the well-studied surface plasmon
lifetime. \cite{kreibig, kawab66_JPSJ, barma89_JPCM, yanno92_AP, molin02_PRB,
weick05_PRB, weick06_PRB} For experimentally available magnetic fields, the 
surface magnetoplasmon linewidths are of the same order as the linewidth $\gamma$ of the 
usual surface plasmon. Since typically $\omega_\mathrm{c}\ll\gamma$, 
it is very difficult to individually resolve the collective magnetoplasmon
resonances in a direct absorption experiment.
However, we propose an indirect way to detect the predicted magnetic-field-induced modification of the absorption spectrum
by means of optical spectroscopy experiments. 
When the electric field of the exciting laser is linearly polarized parallel to
the external magnetic field, only the usual surface plasmon is triggered.
In contrast, when the electric field is perpendicular to the external magnetic
field, only the
surface magnetoplasmons are excited (see Fig.~\ref{fig:sketch}). 
Using circularly polarized light, the two surface magnetoplasmon modes can be
individually addressed. 
The difference in the resulting absorption
spectra should be clearly detectable with 
experimentally available magnetic fields.

Furthermore, we argue that our results should be applicable, at least
qualitatively, for
ferro- or superparamagnetic nanoparticles, where the saturation magnetization is
similar to an effective external magnetic field. \cite{kitte63_PRL}
It might therefore be necessary
to include the effect of the magnetic
field on the electronic center-of-mass motion when one attempts to model the effect of
laser light on the magnetization of such nanoparticles. 

The paper is organized as follows: In Sec.~\ref{sec:model}, we present our model
and the relevant collective coordinates to describe the surface magnetoplasmon 
dynamics. We calculate their lifetimes which constitute the main result of the
present paper in Sec.~\ref{sec:lifetimes}. We suggest 
in Sec.~\ref{sec:experiment} optical absorption spectroscopy experiments to
observe the surface magnetoplasmon collective modes
before we conclude in Sec.~\ref{sec:conclusions}. 
Several technical issues are explained in the appendices.

\section{Microscopic modeling of surface magnetoplasmons}
\label{sec:model}
We consider a spherical nanoparticle in vacuum containing $N$ valence electrons of charge
$-e$ and mass $m_\mathrm{e}$. The nanoparticle of radius $a=r_\mathrm{s}N^{1/3}$
with $r_\mathrm{s}=(3/4\pi n_\mathrm{e})^{1/3}$ the Wigner-Seitz radius
($n_\mathrm{e}$ is the electronic density) is subject to a homogeneous magnetic
field $\mathbf{B}=\nabla\times\mathbf{A}$, where $\mathbf{A}$ is the vector potential.
Within the jellium approximation, which assumes a homogeneous
positive background of charge $+Ne$ (Ref.~\onlinecite{brack93_RMP}), the 
electronic Hamiltonian reads \cite{footnote:cgs}
\begin{align}
\label{eq:Hstart}
H=\,&\sum_{i=1}^N
\left[
\frac{1}{2m_\mathrm{e}}\left(\mathbf{p}_i+\frac{e}{c}\mathbf{A}(\mathbf{r}_i)\right)^2
+U(r_i)+\frac{g_\mathrm{e}\mu_\mathrm{B}}{\hbar}\mathbf{B}\cdot\mathbf{s}_i
\right]
\nonumber\\
&+\sum_{\substack{i,j=1\\(i\neq j)}}^N
\frac{e^2}{2|\mathbf{r}_i-\mathbf{r}_j|},
\end{align}
with $g_\mathrm{e}\simeq2$ the electronic g-factor, $\mu_\mathrm{B}=e\hbar/2m_\mathrm{e}c$
the Bohr magneton, $c$ the
speed of light, and $\mathbf{s}$ the spin operator. The
single-particle confinement 
\begin{equation}
\label{eq:U}
U(r)=\frac{Ne^2}{2a^3}\left(r^2-3a^2\right)\Theta(a-r)-\frac{Ne^2}{r}\Theta(r-a)
\end{equation}
is harmonic with the Mie frequency
$\omega_\mathrm{M}=\sqrt{Ne^2/m_\mathrm{e}a^3}$ inside the particle and
Coulomb-like outside. In Eq.~\eqref{eq:U}, $\Theta(x)$ denotes the Heaviside step function. 

Assuming that the magnetic field points in the $z$ direction,
$\mathbf{B}=B\mathbf{e}_z$ (see Fig.~\ref{fig:sketch}), 
we can express the Hamiltonian \eqref{eq:Hstart} in
the symmetric gauge $\mathbf{A}=(-y, x, 0)B/2$ as
\begin{equation}
\label{eq:H}
H=\sum_{i}
\left[
\frac{p_i^2}{2m_\mathrm{e}}+U(r_i)
+\frac{\omega_\mathrm{c}}{2}\left(l_{z, i}+2s_{z, i}\right)
\right]
+\sum_{i,j}
\frac{e^2}{2|\mathbf{r}_i-\mathbf{r}_j|},
\end{equation}
with $\omega_\mathrm{c}=eB/m_\mathrm{e}c$ the cyclotron frequency, and $l_z$ and
$s_z$ the $z$-components of angular momentum and spin, respectively. 
Notice that in writing Eq.~\eqref{eq:H}, we omitted the diamagnetic term 
$\sum_im_\mathrm{e}\omega_\mathrm{c}^2(x_i^2+y_i^2)/8$ which
is quadratic in the magnetic field and of the order of
$Nm_\mathrm{e}\omega_\mathrm{c}^2a^2$. 
As the Mie frequency typically lies in the optical range,
and $\hbar\omega_\mathrm{c}/B=0.12\,\mathrm{meV/T}$, we have 
$\omega_\mathrm{c}\ll\omega_\mathrm{M}$, even for the highest presently
achievable static magnetic fields of several tens of teslas. 
Thus, the diamagnetic term represents a small correction to
the single-particle confinement \eqref{eq:U}, which is proportional to
$(\omega_\mathrm{c}/\omega_\mathrm{M})^2$. 
In the sequel we work to first order in the small parameter
$\omega_\mathrm{c}/\omega_\mathrm{M}$.

\subsection{Separation into collective and relative coordinates}
A monochromatic electric field of (complex) amplitude $\mathbf{E}_0$ and frequency
$\omega$ can be used to excite the electronic system. In the long-wavelength
limit, the coupling to such an electric field is described by the Hamiltonian
\begin{equation}
\label{eq:H_f}
H_\mathrm{f}=-e\sum_{i}\mathbf{r}_i\cdot\mathbf{E}_0\,\mathrm{e}^{\mathrm{i}\omega t}.
\end{equation}
This field only couples to the
electronic center of mass. Hence, it is appealing to decompose the Hamiltonian
\eqref{eq:H} by introducing the
electronic center-of-mass coordinate $\mathbf{R}=\sum_i\mathbf{r}_i/N$ and its
conjugate momentum $\mathbf{P}=\sum_i\mathbf{p}_i$. \cite{gerchi02_PRA, weick06_PRB} 
The relative coordinates are
denoted by $\mathbf{r}'_i=\mathbf{r}_i-\mathbf{R}$ and
$\mathbf{p}'_i=\mathbf{p}_i-\mathbf{P}/N$. Introducing this new set of
coordinates in Eq.~\eqref{eq:H} and assuming that the displacement of
the center of mass is much smaller than the nanoparticle size ($|\mathbf{R}|\ll
a$), we obtain the decomposition
\begin{equation}
\label{eq:H_decompo}
H=H_\mathrm{cm}+H_\mathrm{rel}+H_\mathrm{c}+H_\mathrm{Z}, 
\end{equation}
where $H_\mathrm{cm}$ and $H_\mathrm{rel}$ are the center-of-mass and
relative-coordinate Hamiltonians, respectively, and $H_\mathrm{c}$ is the coupling
between them. The Zeeman term
\begin{equation}
\label{eq:H_Z}
H_\mathrm{Z}=\omega_\mathrm{c}\sum_is_{z,i}
\end{equation}
accounts for the spin degrees of freedom.
The decomposition \eqref{eq:H_decompo} is reminiscent of the well-studied case
where the degree of freedom of interest (in our case, the electronic center of
mass) is coupled to a large reservoir or environment (the relative coordinates).
\cite{weiss}
The interaction with the reservoir leads to the dissipation of the collective
coordinate energy. The number of degrees of freedom of the reservoir is proportional to the
number of electrons in the nanoparticle, and thus, it is not very large. However,
it has been shown \cite{seoan07_EPJD, weick07_EPJD} that such a reservoir is sufficient to
constitute a well-defined environment for the collective excitation, provided
the nanoparticle is not extremely small. 

The center-of-mass Hamiltonian appearing in Eq.~\eqref{eq:H_decompo} can be decomposed in two parts as
\begin{equation}
\label{eq:H_cm}
H_\mathrm{cm}=H_\mathrm{cm}^\parallel+H_\mathrm{cm}^\perp,
\end{equation}
with the part depending on the collective coordinate $Z$ parallel to the
magnetic field
\begin{equation}
\label{eq:H_cm_Z}
H_\mathrm{cm}^\parallel=\frac{P_Z^2}{2M}
+\frac{M\tilde\omega_\mathrm{M}^2}{2}Z^2,
\end{equation}
and the transverse component
\begin{equation}
\label{eq:H_cm_XY}
H_\mathrm{cm}^\perp=\frac{P_X^2+P_Y^2}{2M}
+\frac{M\tilde\omega_\mathrm{M}^2}{2}\left(X^2+Y^2\right)
+\frac{\omega_\mathrm{c}}{2}L_Z.
\end{equation}
Here $M=Nm_\mathrm{e}$ and $L_Z=XP_Y-YP_X$.
The frequency
$\tilde\omega_\mathrm{M}=\omega_\mathrm{M}\sqrt{1-N_\mathrm{out}/N}$ is slightly
redshifted as compared to the bare Mie frequency by the spill-out effect,
\cite{brack93_RMP, weick06_PRB, bertsch, kreibig} where $N_\mathrm{out}$ is the 
number of electrons outside the nanoparticle. In what follows, we 
approximate $\tilde\omega_\mathrm{M}$ by $\omega_\mathrm{M}$ for simplicity. The
Hamiltonian for the relative
coordinates reads
\begin{equation}
\label{eq:H_rel}
H_\mathrm{rel}=\sum_{i}
\left[
\frac{{p'_i}^2}{2m_\mathrm{e}}+U(r'_i)
+\frac{\omega_\mathrm{c}}{2}l'_{z, i}
\right]
+\sum_{{i,j}}
\frac{e^2}{2|\mathbf{r}'_i-\mathbf{r}'_j|},
\end{equation}
and the coupling Hamiltonian
\begin{equation}
\label{eq:H_c}
H_\mathrm{c}=\sum_i\mathbf{R}\cdot\left[\nabla
U(r'_i)\right]\Big|_{\mathbf{R}=0},
\end{equation}
with 
$\mathbf{R}\cdot\nabla
U(r)=m_\mathrm{e}\omega_\mathrm{M}^2\mathbf{R}\cdot\mathbf{r}f(r)$
and 
\begin{equation}
\label{eq:f}
f(r)=
\Theta(a-r)+\left(\frac{a}{r}\right)^3\Theta(r-a).
\end{equation}

The parallel part \eqref{eq:H_cm_Z} of the center-of-mass Hamiltonian describes a harmonic oscillator which is 
independent of the magnetic field, and whose classical motion is sketched in
Fig.~\ref{fig:sketch}.
Its lowest excitation corresponds to the usual surface plasmon. \cite{weick05_PRB,
weick06_PRB} In contrast, the transverse
part \eqref{eq:H_cm_XY} includes the effect of
the magnetic field. 
The two associated collective modes are called surface magnetoplasmons. Their
corresponding classical orbits are sketched in Fig.~\ref{fig:sketch}.
Thus, an electric field which is polarized parallel to the magnetic field will
excite only the usual surface plasmon in the $z$ direction. In contrast, when
$\mathbf{E}_0\perp\mathbf{B}$, only the surface magnetoplasmons will be
excited. In the general case, when $\mathbf{E}_0$ has components perpendicular
and parallel to $\mathbf{B}$, both the surface plasmon and the surface
magnetoplasmons will be excited. Therefore, the relative orientation of the
laser polarization and the external magnetic field allow one to
selectively excite the different collective modes. 

The Hamiltonian \eqref{eq:H_cm_XY} is similar to the one encountered in the
context of quasi-two-dimensional semiconductor quantum dots. \cite{jacak, reima02_RMP} The
main difference is that, unlike Eq.~\eqref{eq:U}, the single-particle
confinement in quantum dots is well approximated by a harmonic potential for all
relevant $r$. Hence, due to Kohn's theorem, \cite{jacak, reima02_RMP, kohn61_PR} 
center-of-mass and relative
coordinates decouple. In metallic nanoparticles, the Coulomb tail of the single-particle
confinement \eqref{eq:U} leads to a non-negligible coupling Hamiltonian
\eqref{eq:H_c} and therefore to the decay of the surface (magneto)plasmon
excitations. Another difference is that, in quasi-two-dimensional semiconductor
quantum dots, the relatively small effective electronic mass renders the cyclotron
frequency of the order of the confining one, such that the diamagnetic term
in the Hamiltonian \eqref{eq:Hstart} cannot be omitted in the context of quantum
dots.

\subsection{Effect of the magnetic field on the center-of-mass oscillation}
The Hamiltonian \eqref{eq:H_cm_Z} can be written as 
\begin{equation}
H_\mathrm{cm}^\parallel=\hbar\omega_\mathrm{M}\left(b^\dagger b+\frac 12\right)
\end{equation}
in terms of the bosonic operators
\begin{subequations}
\label{eq:b}
\begin{align}
b&=\frac{1}{\sqrt{2}}\left(\frac{Z}{\ell_\mathrm{osc}}+\mathrm{i}\frac{P_Z\ell_\mathrm{osc}}{\hbar}\right),\\
b^\dagger&=\frac{1}{\sqrt{2}}\left(\frac{Z}{\ell_\mathrm{osc}}-\mathrm{i}\frac{P_Z\ell_\mathrm{osc}}{\hbar}\right),
\end{align}
\end{subequations}
with $\ell_\mathrm{osc}=\sqrt{\hbar/M\omega_\mathrm{M}}$ being the oscillator 
length. 

The Hamiltonian $H_\mathrm{cm}^{\perp}$ of Eq.~\eqref{eq:H_cm_XY} can be diagonalized by means of 
Fock-Darwin states. \cite{jacak} Introducing the new variable 
$\xi=(X+\mathrm{i}Y)/\sqrt{2}$ and its complex conjugate and the bosonic operators
\begin{subequations}
\label{eq:b_pm}
\begin{align}
b_+&=\frac{1}{\sqrt{2}}\left(\frac{\xi^*}{\ell_\mathrm{osc}}
+\ell_\mathrm{osc}\frac{\partial}{\partial\xi}\right), \\
b_+^\dagger&=\frac{1}{\sqrt{2}}\left(\frac{\xi}{\ell_\mathrm{osc}}
-\ell_\mathrm{osc}\frac{\partial}{\partial\xi^*}\right), \\
b_-&=\frac{1}{\sqrt{2}}\left(\frac{\xi}{\ell_\mathrm{osc}}
+\ell_\mathrm{osc}\frac{\partial}{\partial\xi^*}\right),\\ 
b_-^\dagger&=\frac{1}{\sqrt{2}}\left(\frac{\xi^*}{\ell_\mathrm{osc}}
-\ell_\mathrm{osc}\frac{\partial}{\partial\xi}\right),
\end{align}
\end{subequations}
we can write
\begin{equation}
H_\mathrm{cm}^\perp=\hbar\omega_+\left(b_+^\dagger b^{\phantom{\dagger}}_++\frac 12\right)
+\hbar\omega_-\left(b_-^\dagger b^{\phantom{\dagger}}_-+\frac 12\right).
\end{equation}
The frequencies of the two magnetoplasmon excitations read \cite{jacak}
\begin{equation}
\label{eq:omega_pm}
\omega_\pm=\omega_\mathrm{M}
\pm\frac{\omega_\mathrm{c}}{2}.
\end{equation}
The surface magnetoplasmon with the larger (smaller) frequency $\omega_+$
($\omega_-$)  rotates counterclockwise (clockwise) in the plane perpendicular to
the magnetic field (see Fig.~\ref{fig:sketch}). Notice that one has
$\omega_+>\omega_-$ since the Lorentz force $-Ne\mathbf{\dot R}\times\mathbf{B}$
increases (decreases) the strength of the 
confinement seen by the collective mode with frequency $\omega_+$
($\omega_-$).

As is well known, \cite{jacak, reima02_RMP} the effect of the magnetic field is 
thus to split the usual surface plasmon
in the plane perpendicular to the field axis in two surface magnetoplasmon
excitations whose frequencies are separated by the cyclotron frequency
$\omega_\mathrm{c}$. 
However, the experimental observation of the two surface magnetoplasmon modes
will be limited by the linewidths of these collective excitations. We address
this important issue in Sec.~\ref{sec:lifetimes}.

\subsection{Mean-field approximation for the environment}
\label{sec:mean-field}
The Hamiltonian for the relative coordinates \eqref{eq:H_rel} contains the
electron-electron interactions. Assuming that the full correlations are not
crucial for the present problem, we treat the interactions on a mean-field
level.
Numerical calculations using the local density approximation and performed in
the absence of a magnetic field \cite{weick05_PRB} suggest that the
self-consistent potential treating the interactions on a mean-field level 
can be approximated by 
\begin{equation}
\label{eq:V(r)}
V(r)=V_0\Theta(r-a),
\end{equation}
a spherical square well of depth
$V_0=\varepsilon_\mathrm{F}+W$, where $\varepsilon_\mathrm{F}$ and $W$ are the
Fermi energy and the work function of the considered nanoparticle, respectively.
We expect that the square well shape of the effective potential remains a good
approximation for $\omega_\mathrm{c}\ll\omega_\mathrm{M}$. \cite{tanak96_PRB}
We thus write, at mean-field level, 
\begin{equation}
\label{eq:H_MF}
H_\mathrm{rel}=\sum_{i}
\left[
\frac{{p'_i}^2}{2m_\mathrm{e}}+V(r'_i)
+\frac{\omega_\mathrm{c}}{2}l'_{z, i}
\right].
\end{equation}

In what follows, we rewrite the above Hamiltonian together
with the Zeeman Hamiltonian \eqref{eq:H_Z} in
second-quantized form, \cite{footnote:N}
\begin{equation}
\label{eq:H_rel_MF}
H_\mathrm{rel}+H_\mathrm{Z}=\sum_{\alpha\sigma}\varepsilon_{\alpha\sigma} 
c_{\alpha\sigma}^\dagger c^{\phantom{\dagger}}_{\alpha\sigma} ,
\end{equation}
where the operators $c_{\alpha\sigma}^\dagger$ ($c^{\phantom{\dagger}}_{\alpha\sigma}$) 
create (annihilate) one-body
eigenstates $|\alpha,\sigma\rangle$ with eigenenergies
$\varepsilon_{\alpha\sigma}=\varepsilon_\alpha-\sigma\hbar\omega_\mathrm{c}/2$ in the
mean-field potential $V(r)$. Here $\alpha$ is a shorthand notation for the
orbital quantum numbers $(n_\alpha, l_\alpha, m_\alpha)$, with $n_\alpha$,
$l_\alpha$, and $m_\alpha$ being the principal, azimuthal, and magnetic
quantum numbers, respectively,
while $\sigma=+1$ ($-1$) corresponds to spin up (down).
It is important to realize that the Hamiltonians of Eqs.~\eqref{eq:H_MF} and
\eqref{eq:H_rel_MF} do contain the electron-electron interactions (at mean-field
level) and thus that our subsequent results concerning the surface
magnetoplasmon lifetimes incorporate the crucial role played by the electronic 
interactions.

\subsection{Coupling of the center of mass to the environment}
Under our mean-field assumption presented in Sec.~\ref{sec:mean-field}
and using Eqs.~\eqref{eq:b} and \eqref{eq:b_pm}, the
coupling Hamiltonian \eqref{eq:H_c} can be written as
\begin{align}
\label{eq:H_c_2}
H_\mathrm{c}=&\;
m_\mathrm{e}\omega_\mathrm{M}^2\frac{\ell_\mathrm{osc}}{\sqrt{2}}\sum_{\alpha\beta\sigma}
c_{\alpha\sigma}^\dagger c^{}_{\beta\sigma}
\Big[
d_{\alpha\beta}^{\sigma}\left(b^\dagger+b\right)
\nonumber\\
&
+d_{\alpha\beta,+}^{\sigma}\left(b_-^\dagger+b_+\right)
+d_{\alpha\beta,-}^{\sigma}\left(b_+^\dagger+b_-\right)
\Big]
\end{align}
with the matrix elements 
\begin{subequations}
\label{eq:def_ME}
\begin{align}
d_{\alpha\beta}^{\sigma}&=\langle\alpha, \sigma|zf(r)|\beta, \sigma\rangle,\\
d_{\alpha\beta, \pm}^{\sigma}&=\frac{1}{\sqrt{2}}\langle\alpha, \sigma|(x\pm\mathrm{i}y)f(r)|\beta,
\sigma\rangle, 
\end{align}
\end{subequations}
which contain the function $f(r)$ of Eq.~\eqref{eq:f}.
The spherical symmetry of the wave functions associated with the Hamiltonian
$\eqref{eq:H_rel_MF}$ allows us to decompose
the matrix elements \eqref{eq:def_ME} into
angular and radial parts, 
\begin{subequations}
\label{eq:matrix_elements}
\begin{align}
d_{\alpha\beta}^{\sigma}&=\mathcal{A}_{l_\alpha, l_\beta}^{m_\alpha, m_\beta}
\mathcal{R}(\varepsilon_{\alpha\sigma}, \varepsilon_{\beta\sigma}),\\
d_{\alpha\beta, \pm}^{\sigma}&=\mathcal{A}_{l_\alpha, l_\beta, \pm}^{m_\alpha, m_\beta}
\mathcal{R}(\varepsilon_{\alpha\sigma}, \varepsilon_{\beta\sigma}).
\end{align}
\end{subequations}
The angular parts can be expressed in terms of Wigner $3j$ symbols 
\cite{edmonds} as
\begin{subequations}
\label{eq:angular}
\begin{align}
\label{eq:angular_a}
\mathcal{A}_{l_\alpha, l_\beta}^{m_\alpha, m_\beta}=&\;
(-1)^{m_\alpha}\sqrt{(2l_\alpha+1)(2l_\beta+1)}
\nonumber\\
&\;\times
\begin{pmatrix}
l_\alpha & l_\beta & 1\\
0 & 0 & 0
\end{pmatrix}
\begin{pmatrix}
l_\alpha & l_\beta & 1\\
-m_\alpha & m_\beta & 0
\end{pmatrix},
\\
\label{eq:angular_b}
\mathcal{A}_{l_\alpha, l_\beta, \pm}^{m_\alpha, m_\beta}=&\;
\mp(-1)^{m_\alpha}\sqrt{(2l_\alpha+1)(2l_\beta+1)}
\nonumber\\
&\;\times
\begin{pmatrix}
l_\alpha & l_\beta & 1\\
0 & 0 & 0
\end{pmatrix}
\begin{pmatrix}
l_\alpha & l_\beta & 1\\
-m_\alpha & m_\beta & \pm 1
\end{pmatrix},
\end{align}
\end{subequations}
with the selection rules $l_\alpha=l_\beta\pm1$ for Eqs.~\eqref{eq:angular}, and $m_\alpha=m_\beta$ and
$m_\alpha=m_\beta\pm1$ for 
Eqs.~\eqref{eq:angular_a} and \eqref{eq:angular_b}, respectively.
The radial matrix element can be
approximated by
\cite{yanno92_AP}
\begin{equation}
\mathcal{R}(\varepsilon_{\alpha\sigma},
\varepsilon_{\beta\sigma})=\frac{2\hbar^2}{m_\mathrm{e}a}
\frac{\sqrt{\varepsilon_{\alpha\sigma}\varepsilon_{\beta\sigma}}}
{(\varepsilon_{\alpha\sigma}-\varepsilon_{\beta\sigma})^2},
\end{equation}
an expression that is obtained under the assumption of infinite confinement
$V_0\to\infty$ in Eq.~\eqref{eq:V(r)}. \cite{footnote:confinment}

\section{Surface magnetoplasmon decay rates}
\label{sec:lifetimes}

\subsection{Landau damping of the surface magnetoplasmon collective modes}
\label{sec:landau}
For nanoparticles having intermediate sizes of the order of a few nanometers,
the Landau damping \cite{bertsch} is the dominating decay channel for the
collective plasmon modes. \cite{footnote:radiation} The surface (magneto)plasmons decay by producing
particle-hole pairs in the electronic environment whose energies correspond to
the ones of the collective excitations.
Treating the coupling Hamiltonian \eqref{eq:H_c_2} as a perturbation, the
corresponding surface plasmon and surface magnetoplasmon decay rates, whose
inverses yield the lifetimes of these collective excitations, can be obtained
from Fermi's golden rule. Assuming zero temperature, \cite{footnote:temperature} 
we have the decay rates 
\cite{weick05_PRB, weick06_PRB}
\begin{equation}
\gamma=\frac{\pi}{\hbar}\left(m_\mathrm{e}\ell_\mathrm{osc}\omega_\mathrm{M}^2\right)^2
\sum_{ph\sigma}{|d_{ph}^{\sigma}|}^2
\delta\left(\hbar\omega_\mathrm{M}-\varepsilon_{p\sigma}+\varepsilon_{h\sigma}\right)
\end{equation}
and 
\begin{equation}
\label{eq:FGR}
\gamma_\pm=\frac{\pi}{\hbar}
\left(m_\mathrm{e}\ell_\mathrm{osc}\omega_\mathrm{M}^2\right)^2
\sum_{ph\sigma}{|d_{ph, \pm}^{\sigma}|}^2\delta
\left(\hbar\omega_\pm-\varepsilon_{p\sigma}+\varepsilon_{h\sigma}\right)
\end{equation}
for the surface plasmon and surface magnetoplasmons, respectively. Here, $p$ and
$h$ denote particle and hole states with energies
$\varepsilon_{p\sigma}>\varepsilon_\mathrm{F}$ and
$\varepsilon_{h\sigma}<\varepsilon_\mathrm{F}$.

Using the expressions \eqref{eq:matrix_elements} for the coupling matrix
elements and the appropriate selection rules that are contained in
Eq.~\eqref{eq:angular}, we obtain to first order in
$\omega_\mathrm{c}/\omega_\mathrm{M}\ll1$
\begin{subequations}
\label{eq:gammas}
\begin{align}
\gamma&=\frac{2\pi}{\hbar}\left(\frac{2\ell_\mathrm{osc}}{a}\right)^2F,\\
\gamma_\pm&=\frac{2\pi}{\hbar}
\left(\frac{2\ell_\mathrm{osc}}{a}\right)^2\left(1\mp\frac{2\omega_\mathrm{c}}{\omega_\mathrm{M}}\right)
F_\pm,
\end{align}
\end{subequations}
where 
\begin{subequations}
\label{eq:F}
\begin{align}
F=
&\int_{\max{(\varepsilon_\mathrm{F},\hbar\omega_\mathrm{M})}}^{\varepsilon_\mathrm{F}
+\hbar\omega_\mathrm{M}}\mathrm{d}\varepsilon\, \varepsilon(\varepsilon-\hbar\omega_\mathrm{M})
\sum_{l,m}\varrho_{l,m}(\varepsilon) \nonumber\\
&\times 
\left[
\left(\mathcal{A}_{l, l+1}^{m,m}\right)^2\varrho_{l+1,m}(\varepsilon-\hbar\omega_\mathrm{M})
\right.\nonumber\\
&
+
\left.
\left(\mathcal{A}_{l, l-1}^{m,m}\right)^2\varrho_{l-1,m}(\varepsilon-\hbar\omega_\mathrm{M})
\right],\\
F_\pm=
&\int_{\max{(\varepsilon_\mathrm{F},\hbar\omega_\pm)}}^{\varepsilon_\mathrm{F}
+\hbar\omega_\pm}\mathrm{d}\varepsilon\, \varepsilon(\varepsilon-\hbar\omega_\pm)
\sum_{l,m}\varrho_{l,m}(\varepsilon) \nonumber\\
&\times 
\left[
\left(\mathcal{A}_{l, l+1, \pm}^{m,m\mp1}\right)^2\varrho_{l+1,m\mp1}(\varepsilon-\hbar\omega_\pm)
\right.\nonumber\\
&+\left.\left(\mathcal{A}_{l, l-1, \pm}^{m,m\mp1}\right)^2\varrho_{l-1,m\mp1}(\varepsilon-\hbar\omega_\pm)
\right].
\end{align}
\end{subequations}
Here $\varrho_{l, m}(\varepsilon)$ is the density of states for fixed angular
momentum $l$ and magnetic quantum number $m$. \cite{molin02_PRB} For metals the Fermi wave vector 
$k_\mathrm{F}\simeq\unit[10]{nm^{-1}}$, such that $k_\mathrm{F}a\gg1$ for nanoparticles having a 
radius of more than a nanometer. Thus, we can resort to 
the semiclassical approximation \cite{gutzwiller, brack} to evaluate the 
density of states. To leading order in $\hbar$, we have
\cite{weick05_PRB, footnote:oscillations}
\begin{equation}
\label{eq:DOS}
\varrho_{l, m}(\varepsilon)=
\frac{\sqrt{2m_\mathrm{e}a^2(\varepsilon-\hbar\omega_\mathrm{c}m/2)/\hbar^2-(l+1/2)^2}}
{2\pi(\varepsilon-\hbar\omega_\mathrm{c}m/2)},
\end{equation}
which allows to evaluate Eq.~\eqref{eq:F} in the
semiclassical limit (for details, see Appendix~\ref{app:details}), leading to
\begin{equation}
\label{eq:gamma}
\gamma=\frac{3\varv_\mathrm{F}}{4a}g(\nu)
\end{equation}
for the surface plasmon decay rate.
Here,
$\varv_\mathrm{F}$ denotes the Fermi velocity, and
$\nu=\hbar\omega_\mathrm{M}/\varepsilon_\mathrm{F}$.
The function $g(\nu)$ given in
Appendix~\ref{app:aux} [cf.\ Eq.~\eqref{eq:g}] is a monotonously decreasing
function. It is remarkable that the result of
Eq.~\eqref{eq:gamma}, which has been obtained to first order in the small
parameter $\omega_\mathrm{c}/\omega_\mathrm{M}$, does not present any magnetic field dependence. 
We can thus conclude that for realistic field strengths the surface plasmon is not influenced by the external
magnetic field. 
The result of Eq.~\eqref{eq:gamma} and its well-known $1/a$ size dependence has
been first obtained by Kawabata and Kubo \cite{kawab66_JPSJ} and refined later
by many authors. \cite{barma89_JPCM, yanno92_AP, kreibig, molin02_PRB,
weick05_PRB, weick06_PRB} 

For the two surface magnetoplasmon collective modes, we find
for
$\omega_\mathrm{c}\ll(\omega_\mathrm{M},\varepsilon_\mathrm{F}/\hbar)$
(see Appendix~\ref{app:details} for details) the decay rates 
\begin{equation}
\label{eq:gamma_pm}
\gamma_\pm=\frac{3\varv_\mathrm{F}}{4a}
\left[
g(\nu)\mp\frac{\omega_\mathrm{c}}{\omega_\mathrm{M}}
\left(
\frac{3}{2}g(\nu)
-\frac{\nu}{2}
g'(\nu)
-h(\nu)
\right)
\right]
\end{equation}
which constitute the main results of our work.
Here $g'(\nu)$ denotes the derivative of the function $g(\nu)$ given in Eq.~\eqref{eq:g}, and the function
$h(\nu)$ is given in Appendix~\ref{app:aux} [cf.\ Eq.~\eqref{eq:h}].
The surface magnetoplasmon decay rates \eqref{eq:gamma_pm} 
present the same size dependence as that of the surface
plasmon, Eq.~\eqref{eq:gamma}, and obviously verify $\gamma_\pm=\gamma$ for 
zero magnetic field ($\omega_\mathrm{c}=0$).
It is important to realize that the function $3g(\nu)/2-\nu g'(\nu)/2-h(\nu)$
entering the result of Eq.~\eqref{eq:gamma_pm}
is positive for $\nu\lesssim4.51$, i.e., it is positive for any realistic values
of the parameter $\nu=\hbar\omega_\mathrm{M}/\varepsilon_\mathrm{F}$.
Thus, we conclude that $\gamma_+$ ($\gamma_-$)
decreases (increases) linearly for increasing $\omega_\mathrm{c}$ (i.e., increasing
magnetic field). This is due to the fact that the frequency $\omega_+$ ($\omega_-$) of the
corresponding mode shows the opposite behavior, i.e., it increases (decreases)
with $\omega_\mathrm{c}$ [cf.\ Eq.~\eqref{eq:omega_pm}]. The behavior of the
decay rates $\gamma_\pm$ of Eq.~\eqref{eq:gamma_pm} as a function of the
magnetic field can
easily be explained as follows: The typical dipole matrix elements between
particle and hole states separated by an energy 
$\Delta\varepsilon_{ph}=\varepsilon_p-\varepsilon_h$
entering the Fermi golden rule \eqref{eq:FGR} scale as $d^\sigma_{ph,\pm}\sim
1/\Delta\varepsilon_{ph}^2$, while the density of particle-hole states
fulfilling the selection rules dictated by the angular part \eqref{eq:angular_b} of the matrix
elements $d_{ph,\pm}^\sigma$ is linear
in $\Delta\varepsilon_{ph}$. Since energy conservation requires that 
$\Delta\varepsilon_{ph}=\hbar\omega_\pm$, this argument, valid
in the limit 
$\hbar\omega_\mathrm{c}\ll\hbar\omega_\mathrm{M}\ll\varepsilon_\mathrm{F}$
(for details, see Ref.~\onlinecite{seoan07_EPJD}),
where $g(0)=1$, $g'(0)=0$, and $h(0)=0$, 
yields the scaling $\gamma_\pm\sim
1/\omega_\pm^3\sim1\mp3\omega_\mathrm{c}/2\omega_\mathrm{M}$,
consistent with Eq.~\eqref{eq:gamma_pm}.

\subsection{Extension to dielectric environments and noble-metal and ferromagnetic nanoparticles}
In the case of alkali clusters in vacuum, the result of Eq.~\eqref{eq:gamma} for
the surface plasmon decay rate agrees quantitatively with numerical calculations
using the time-dependent local density approximation as well as with
experiments.
We take this as a strong
indication that the surface magnetoplasmon decay rate \eqref{eq:gamma_pm} is
quantitatively valid for alkaline clusters in vacuum as well. 
When one considers alkaline nanoparticles in a dielectric environment or 
noble-metal nanoparticles (in vacuum or in an embedding matrix), one has to take into account
the steepness of the effective mean-field potential $V(r)$ 
entering Eq.~\eqref{eq:H_MF} in the derivation of
Eq.~\eqref{eq:gamma} to get quantitative agreement with numerical calculations
as well as with experiments. \cite{weick05_PRB} 
The resulting linewidth of the surface plasmon excitation decreases as the dielectric
constants of the metal and of the dielectric environment increase. This is also
the case for the linewidths of the surface magnetoplasmons.
Hence, their observation will be facilitated by a dielectric environment.

Furthermore, our results for 
the decay rates of the collective electronic excitations should, at least 
qualitatively, be applicable to ferromagnetic nanoparticles, provided one adds 
to the external magnetic field
$\mathbf{B}$ the internal magnetic field $4\pi\mathbf{M}_\mathrm{s}$,
\cite{kitte63_PRL} with $\mathbf{M}_\mathrm{s}$ being the saturation magnetization.
This internal field couples to the orbital degrees of freedom and is of the 
order of \unit[2]{T} for bulk transition ferromagnets.
Thus, in saturated ferromagnetic nanoparticles, the surface magnetoplasmons might exist
even in the absence of an external magnetic field.
This possible extension of our considerations toward ferromagnetic
nanoparticles might be important for the analysis of the magnetization dynamics in such
systems \cite{andra06_PRL, kiril10_RMP} since the surface magnetoplasmons create 
electromagnetic fields inside the particle that might affect the magnetic 
moments responsible for the magnetism in these nanoparticles.

\section{Experimental detection of surface magnetoplasmons}
\label{sec:experiment}

\subsection{Absorption profiles}
\label{sec:profile}
For the nanoparticle radii of a few nanometer that we are considering, the
extinction spectrum is dominated by absorption. \cite{kreibig, born} Assuming a
Lorentzian profile for the line shapes of the collective modes, the (normalized)
absorption cross section of photons with frequency $\omega$ is given by
\begin{equation}
\label{eq:sigma_parallel}
\sigma_\parallel(\omega)=\frac{\gamma/2\pi}{(\omega-\omega_\mathrm{M}
)^2+(\gamma/2)^2}
\end{equation}
when the illuminating electric field is polarized parallel to the magnetic field
[$\mathbf{E}_0=E_0\mathbf{e}_z\parallel\mathbf{B}$, cf.\ Eq.~\eqref{eq:H_f}].
In that case, only the usual magnetic-field independent surface
plasmon is excited. 

\begin{figure}[tb]
\includegraphics[width=\linewidth]{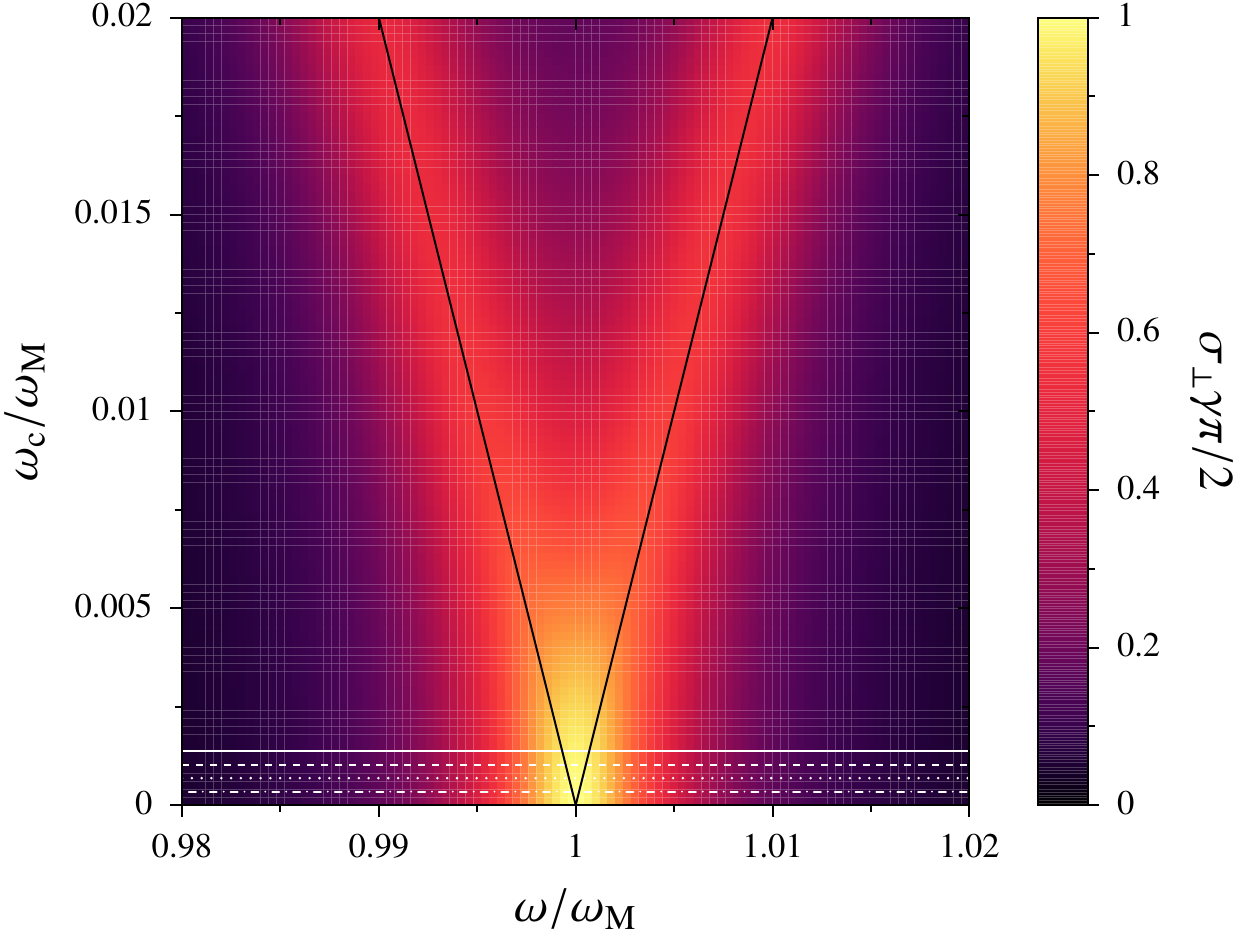}
\caption{\label{fig:absorption2D}%
(Color online) Absorption cross section $\sigma_\perp$ for linearly polarized
light with electric
field perpendicular to the external magnetic field [Eq.~\eqref{eq:sigma_perp}]
as a function of the photon frequency $\omega$ and the cyclotron frequency
$\omega_\mathrm{c}$, both scaled with the Mie frequency $\omega_\mathrm{M}$. 
The chosen parameters correspond to a sodium nanoparticle
($r_\mathrm{s}=3.93a_0$, where $a_0$ is the Bohr radius) of radius $a=\unit[10]{nm}$, having
$\hbar\omega_\mathrm{M}=\unit[3.5]{eV}$. 
The horizontal white lines correspond to magnetic
fields $B=\unit[10]{T}$ (dash-dotted line), $B=\unit[20]{T}$ (dotted
line), $B=\unit[30]{T}$ (dashed line), and $B=\unit[40]{T}$ (solid line).
The black lines indicate the surface magnetoplasmon resonance frequencies,
$\omega_-$ (left line) and $\omega_+$ (right line) [see
Eq.~\eqref{eq:omega_pm}].}
\end{figure}

In the case where $\mathbf{E}_0\perp\mathbf{B}$, the
two collective surface magnetoplasmon modes are triggered. 
Using linearly polarized light with
$\mathbf{E}_0=E_0(\cos{\varphi}\,\mathbf{e}_x+\sin{\varphi}\,\mathbf{e}_y)$,
both magnetoplasmons are excited, and the absorption cross section reads
\begin{equation}
\label{eq:sigma_perp}
\sigma_\perp(\omega)=\frac{1}{2}\left[\sigma_+(\omega)+\sigma_-(\omega)\right], 
\end{equation}
with
\begin{equation}
\label{eq:sigma_plus}
\sigma_+(\omega)=\frac{\gamma_+/2\pi}{(\omega-\omega_+)^2+(\gamma_+/2)^2}
\end{equation}
and 
\begin{equation}
\label{eq:sigma_minus}
\sigma_-(\omega)=\frac{\gamma_-/2\pi}{(\omega-\omega_-)^2+(\gamma_-/2)^2}.
\end{equation}
The absorption profile $\sigma_\perp$, scaled with the maximum
$2/\pi\gamma$ of $\sigma_\parallel$, is shown in Fig.~\ref{fig:absorption2D}
as a function of the photon frequency $\omega$ and for increasing cyclotron
frequency $\omega_\mathrm{c}$ for the case of a sodium nanoparticle of radius
$a=\unit[10]{nm}$ having a collective surface plasmon resonance at
$\hbar\omega_\mathrm{M}=\unit[3.5]{eV}$.
In the regime of currently available static magnetic fields, up to about \unit[40]{T}
(horizontal white solid line in Fig.~\ref{fig:absorption2D}),
the two magnetoplasmon modes (indicated by black lines in
Fig.~\ref{fig:absorption2D}) are separated by a frequency $\omega_\mathrm{c}$
which is smaller than the surface magnetoplasmon linewidths $\gamma_+$ and
$\gamma_-$ [Eq.~\eqref{eq:gamma_pm}]. These linewidths are of the same order as that
of the surface plasmon $\gamma$ [Eq.~\eqref{eq:gamma}]. \cite{footnote:gamma}
Hence, the linewidths are larger than the separation of the
resonances such that the two modes may not be directly resolved in an
absorption spectrum. \cite{footnote:observability}
It should be noted that even using pulsed magnetic fields that can reach up to
\unit[70]{T} would not allow one to clearly separate the two modes.
As can be seen from Fig.~\ref{fig:absorption2D}, only for unrealistically large 
$\omega_\mathrm{c}/\omega_\mathrm{M}\gtrsim \unit[1]{\%}$ do the two collective
excitations exhibit separate maxima.

A way to individually address the surface magnetoplasmons is to 
use circularly polarized light. In the case where 
$\mathbf{E}_0=E_0(\mathbf{e}_x+\mathrm{i}\mathbf{e}_y)/\sqrt{2}$ 
[$\mathbf{E}_0=E_0(\mathbf{e}_x-\mathrm{i}\mathbf{e}_y)/\sqrt{2}$], one can selectively
excite the surface magnetoplasmon with frequency $\omega_+$ [$\omega_-$],
and the absorption
spectrum is given by Eq.~\eqref{eq:sigma_plus} [Eq.~\eqref{eq:sigma_minus}].
But for currently experimentally achievable magnetic fields, the displacement in
frequency, as well as the modification of the linewidth of these peaks, is so
small that it might not be observable. 

In addition, for an ensemble of
nanoparticles, the inhomogeneous broadening due to their dispersion in size
further increases the total linewidth.
Thus, the direct and unambiguous observation of two distinct resolved peaks in
the absorption spectrum of an ensemble of nanoparticles using linearly polarized
light might not be possible. Moreover, using circularly polarized light which
enables one to
excite a single surface magnetoplasmon mode may not present a sufficient
magnetic-field-dependent behavior in the absorption cross section.
However,
we suggest in the following two differential measurements 
where one could experimentally identify the effect
of the magnetic field on the collective excitations and the presence of
the surface magnetoplasmons, one using linearly polarized light
(Sec.~\ref{sec:linear}) and the other, which is more efficient, using circularly polarized
light (Sec.~\ref{sec:circular}). The effect of inhomogeneous broadening on these
proposed differential measurements is
discussed in Sec.~\ref{sec:inhomogeneous}.

\subsection{Detection with linearly polarized light}
\label{sec:linear}
\begin{figure}[tb]
\includegraphics[width=\linewidth]{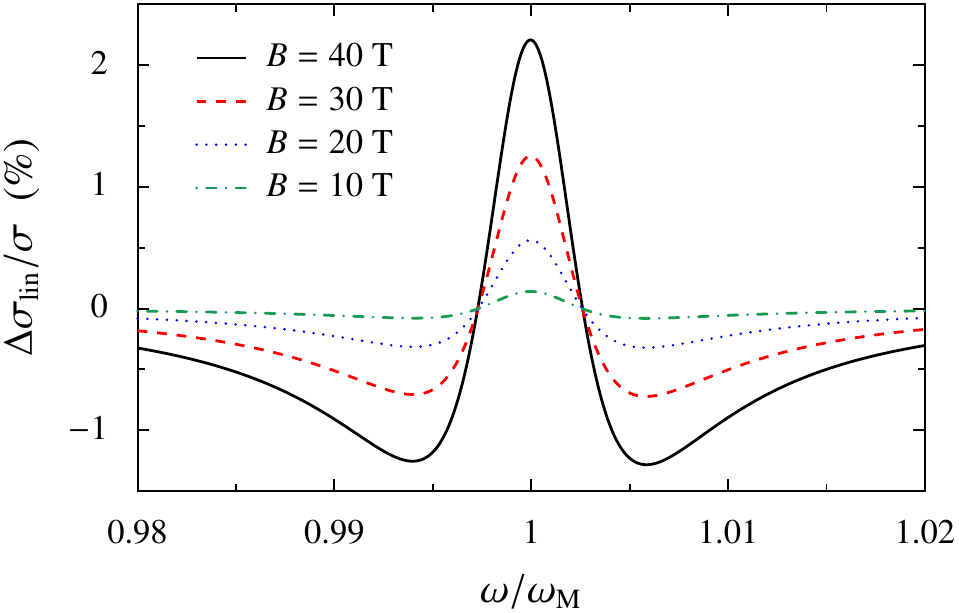}
\caption{\label{fig:absorption}%
(Color online)
Differential absorption cross section $\Delta\sigma_\mathrm{lin}/\sigma$ of
Eq.~\eqref{eq:sigma_diff} as a function of the photon frequency $\omega$. 
The parameters are the same as in Fig.~\ref{fig:absorption2D}. The values of
the magnetic field used in this figure correspond to the horizontal white
lines in Fig.~\ref{fig:absorption2D}.}
\end{figure}

The relative differential absorption cross section
\begin{equation}
\label{eq:sigma_diff}
\frac{\Delta\sigma_\mathrm{lin}}{\sigma} =
\frac{\sigma_\parallel-\sigma_\perp}{\sigma_\parallel}
\end{equation} 
given by the difference between the absorption cross sections
for the electric field linearly polarized parallel
[Eq.~\eqref{eq:sigma_parallel}] and perpendicular
[Eq.~\eqref{eq:sigma_perp}] to the external magnetic field is depicted in
Fig.~\ref{fig:absorption} for the values of the magnetic field indicated by 
the white lines in Fig.~\ref{fig:absorption2D} (the parameters
are for the same sodium nanoparticles of radius $a=\unit[10]{nm}$). 
It can be seen in Fig.~\ref{fig:absorption} that the differential absorption
$\Delta\sigma_\mathrm{lin}/\sigma$ 
increases for increasing magnetic field strength,
yielding a clear signature of the existence of the surface magnetoplasmon
excitations. The increase of the total linewidth of the absorption resonance
caused by the magnetic field via the modified width and the splitting of the 
surface magnetoplasmons leads to positive values of $\Delta\sigma_\mathrm{lin}/\sigma$
in the center of the resonance and to negative values in the tails. 
In magnetic fields of less than \unit[10]{T}, $\Delta\sigma_\mathrm{lin}/\sigma$ is
very small, at least for the case of sodium nanoparticles.
However, in a field of \unit[40]{T} (solid lines in Figs.~\ref{fig:absorption2D} 
and \ref{fig:absorption}), of the order of the highest static field available in
present-day high magnetic field laboratories, the differential absorption
$\Delta\sigma_\mathrm{lin}/\sigma$ becomes noticeable.
Furthermore, magnetic field pulses of \unit[70]{T} are available with
durations of the order of milliseconds that should be sufficient to measure the
absorption cross section. For the latter value, we expect a relative
differential absorption of about \unit[6]{\%} that should be
detectable (not shown in Fig.~\ref{fig:absorption}).

\subsection{Detection with circularly polarized light}
\label{sec:circular}
\begin{figure}[tb]
\includegraphics[width=\linewidth]{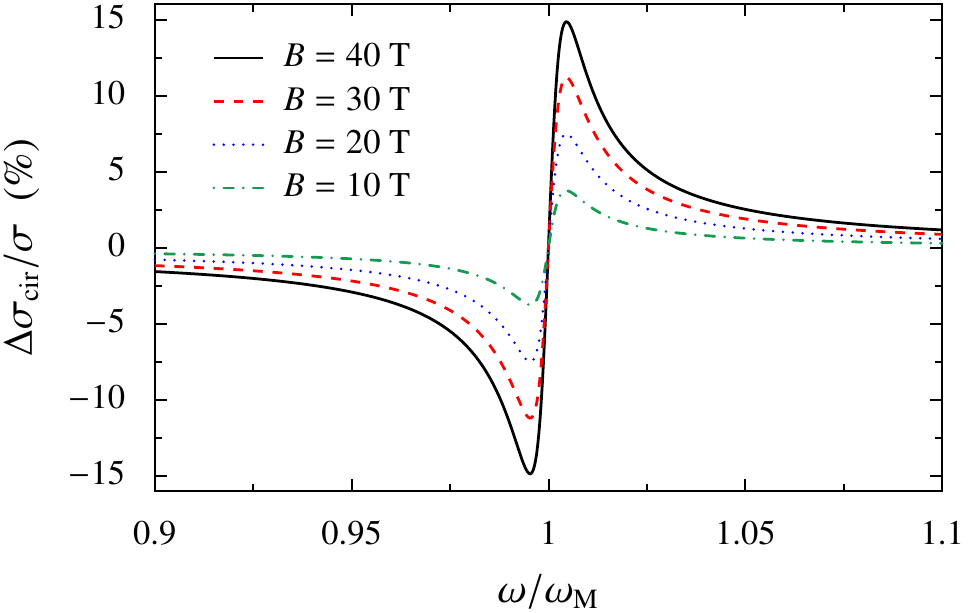}
\caption{\label{fig:absorption_circular}%
(Color online)
Differential absorption cross section $\Delta\sigma_\mathrm{cir}/\sigma$ of
Eq.~\eqref{eq:sigma_cir} as a function of the photon frequency $\omega$. 
The parameters are the same as in Figs.~\ref{fig:absorption2D} and
\ref{fig:absorption}.}
\end{figure}

The relative differential absorption cross section
\begin{equation}
\label{eq:sigma_cir}
\frac{\Delta\sigma_\mathrm{cir}}{\sigma} =
\frac{\sigma_+-\sigma_-}{\sigma_++\sigma_-}
\end{equation}
given by the difference between absorption cross sections for the two circular
polarizations of the electric field, Eqs.~\eqref{eq:sigma_plus} and \eqref{eq:sigma_minus}, is shown in
Fig.~\ref{fig:absorption_circular} for the same parameters as in
Figs.~\ref{fig:absorption2D} and \ref{fig:absorption}. 
It is positive (negative)
for $\omega>\omega_\mathrm{M}$ ($\omega<\omega_\mathrm{M}$) due to the fact that
the resonance frequency $\omega_+$ ($\omega_-$) of the surface magnetoplasmon
mode ``$+$" (``$-$") is larger (smaller) than the surface plasmon frequency $\omega_\mathrm{M}$
[cf.\ Eq.~\eqref{eq:omega_pm}]. As one can see from
Fig.~\ref{fig:absorption_circular}, the differential
absorption cross section $\Delta\sigma_\mathrm{cir}/\Delta\sigma$ of
Eq.~\eqref{eq:sigma_cir} is much more
pronounced than the one of Eq.~\eqref{eq:sigma_diff} using linearly polarized
light (compare with Fig.~\ref{fig:absorption}). Moreover, the frequency range
where $\Delta\sigma_\mathrm{cir}/\sigma$ presents noticeable values is much broader. In the case of circularly
polarized light, $\Delta\sigma_\mathrm{cir}/\sigma$ is already of the order of
several percent at a magnetic field of \unit[10]{T} (green dash-dotted line in
Fig.~\ref{fig:absorption_circular}) and can reach up to about \unit[15]{\%} in a static field
of \unit[40]{T} (black solid line). For a pulsed magnetic field of \unit[70]{T}, we
obtain a maximal relative differential cross section of about \unit[25]{\%} (not
shown in the figure). It thus seems even more efficient to use circularly
polarized light rather than linearly polarized light to 
detect the surface magnetoplasmon modes in metallic nanoparticles.

\subsection{Effect of inhomogeneous broadening}
\label{sec:inhomogeneous}
In Secs.~\ref{sec:profile}, \ref{sec:linear}, and \ref{sec:circular}, we have
implicitly assumed that the nanoparticles irradiated by the laser light do not
present any size dispersion. Although this is a valid approximation for
experiments on single clusters, \cite{klar98_PRL, sonni02_NJP, arbou04_PRL,
berci04_PRL, berci05_NL, dijk05_PRL} most experiments are done on ensembles
of metallic clusters where the inhomogeneous broadening resulting from the size
dependence of the resonance frequency masks the homogeneous linewidth.
\cite{lampr99_APB, stiet00_PRL, bosba02_PRL} 

\begin{figure}[tb]
\includegraphics[width=\linewidth]{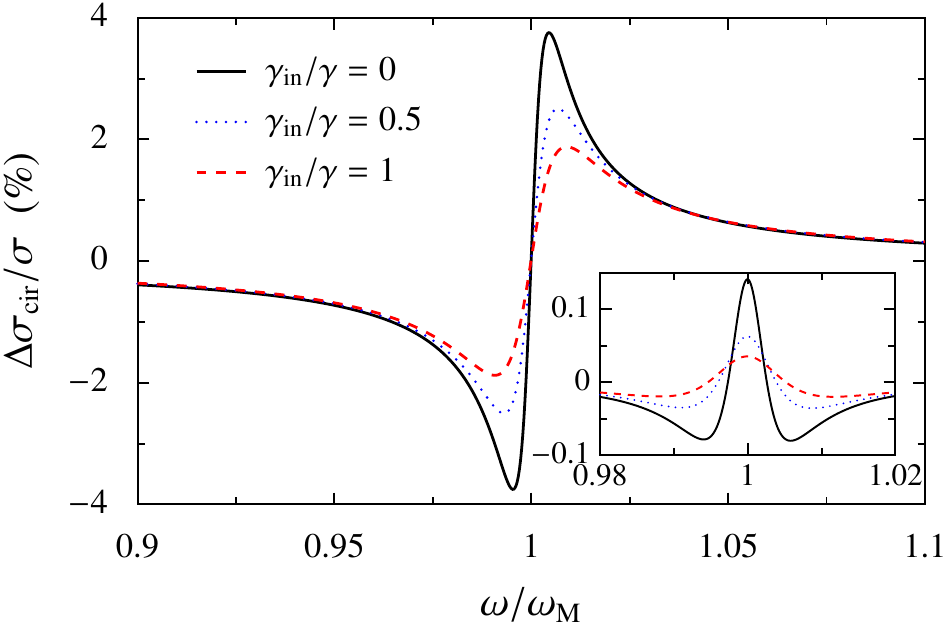}
\caption{\label{fig:absorption_circular_inhomogeneous}%
(Color online)
Differential absorption cross section $\Delta\sigma_\mathrm{cir}/\sigma$ of
Eq.~\eqref{eq:sigma_cir} as a function of the photon frequency $\omega$ in a
magnetic field $B=\unit[10]{T}$ for an ensemble of sodium nanoparticles with
mean radius $\bar a=\unit[10]{nm}$ and for increasing values of the inhomogeneous
linewidth $\gamma_\mathrm{in}$.
The inset is the 
same as the main figure for linearly polarized light, Eq.~\eqref{eq:sigma_diff}.
}
\end{figure}

In this section, we address the
role of inhomogeneous broadening on the experimental
detection of the two surface magnetoplasmon excitations. To this end, 
in the absorption cross sections of Eqs.~\eqref{eq:sigma_parallel},
\eqref{eq:sigma_plus}, and \eqref{eq:sigma_minus}, we
phenomenologically add an inhomogeneous linewidth $\gamma_\mathrm{in}$
resulting from the size dispersion of the ensemble of nanoparticles to the
intrinsic linewidths of the surface (magneto)plasmon collective modes.
The resulting differential
absorption cross sections for sodium nanoparticles with mean radius
$\bar a=\unit[10]{nm}$ in a magnetic field of \unit[10]{T} using 
circularly polarized light [cf.\ Eq.~\eqref{eq:sigma_cir}] are presented in
Fig.~\ref{fig:absorption_circular_inhomogeneous}. Results for the same
parameters using linearly polarized light [cf.\ Eq.~\eqref{eq:sigma_diff}] are
shown in the inset in Fig.~\ref{fig:absorption_circular_inhomogeneous}.

We quantify the inhomogeneous broadening $\gamma_\mathrm{in}$ by its relative
value with respect to
the intrinsic linewidth of the collective modes at zero magnetic field, $\gamma$
[see Eq.~\eqref{eq:gamma}]. Given the size dependence of the resonance frequency
of the surface plasmon excitation which can be estimated from numerical
calculations within the time-dependent local density approximation (see Fig.~7
in Ref.~\onlinecite{weick06_PRB}), and assuming a constant size distribution
with dispersion $\Delta a$ around the mean radius $\bar a$, we find that
$\Delta a/\bar a=\unit[50]{\%}$ ($\Delta a/\bar a=\unit[100]{\%}$) corresponds to
$\gamma_\mathrm{in}/\gamma=0.5$ ($\gamma_\mathrm{in}/\gamma=1$)
for the parameters of Fig.~\ref{fig:absorption_circular_inhomogeneous} (see blue dotted
and red dashed lines).

As one can see from Fig.~\ref{fig:absorption_circular_inhomogeneous}, inhomogeneous broadening tends to weaken the differential 
absorption cross sections. Indeed, the maxima of the differential cross
sections for circularly (Fig.~\ref{fig:absorption_circular_inhomogeneous}) and
linearly polarized light (inset in Fig.~\ref{fig:absorption_circular_inhomogeneous}) 
decrease for increasing inhomogeneous linewidth $\gamma_\mathrm{in}$. For an inhomogeneous
linewidth $\gamma_\mathrm{in}$ comparable to the intrinsic linewidth $\gamma$, the
maximal differential absorption cross section using circularly polarized light,
$\Delta\sigma_\mathrm{cir}/\sigma$, is still of about \unit[2]{\%} in a
magnetic field of \unit[10]{T} (see red dashed line in
Fig.~\ref{fig:absorption_circular_inhomogeneous}). Such 
$\Delta\sigma_\mathrm{cir}/\sigma$ should still be clearly measurable. \cite{stiet00_PRL, bosba02_PRL} 
In the case of linearly polarized light, the maximal
differential absorption cross section is only about \unit[0.03]{\%} for
$\gamma_\mathrm{in}=\gamma$ (see red dashed line in the inset in 
Fig.~\ref{fig:absorption_circular_inhomogeneous}). Such a small value in the difference
between two extinction spectra on ensembles of nanoparticles seems to be
difficult to measure. \cite{stiet00_PRL, bosba02_PRL} 
These results point out that using circularly
polarized light to obtain a clear-cut experimental detection of the two surface
magnetoplasmon excitations in ensembles of metallic nanoparticles presenting a
rather large size (and shape) dispersion is more
appropriate than using linearly polarized light.

\section{Conclusion}
\label{sec:conclusions}
We have analyzed the role of an external magnetic field on collective
excitations in metallic nanoparticles. As is the case in the context of
semiconductor quantum dots, the magnetic field induces two new
resonances in metallic nanoparticles, the surface magnetoplasmons, which can be excited when
the polarization of the electric field has a component perpendicular to the
magnetic field.
Our main result concerns the Landau damping linewidths of these
collective modes that we have calculated. In particular, we have shown treating
the electron-electron interactions within a mean-field approximation
how the magnetic field modifies the
absorption linewidths of the surface magnetoplasmon resonances. In all
realistic cases, the linewidths are much larger than the splitting of the
resonance energies such that one may not resolve them directly in a single
absorption measurement. Nevertheless, if one changes the polarization of the
electric field with respect to the magnetic field from parallel to
perpendicular, a noticeable change in the absorption profiles should be
detectable, at least in very strong magnetic fields, as they are currently available
in high-field laboratories. Using circularly polarized light, which
enables one to selectively excite the surface magnetoplasmon modes, leads to
even larger values of the differential cross section when one changes the
polarization from right-handed to left-handed. 
The proposed differential measurements are expected to allow a detection of the
surface magnetoplasmon modes even in the presence of an inhomogeneous broadening
of the resonances that results from the size and shape dispersion for ensembles
of nanoparticles in matrices.

Our results can be extended to ferromagnetic
nanoparticles, where the internal magnetic field (i.e., the saturation magnetization)
couples to the orbital degrees of freedom.
In addition, when the ground-state magnetization in nanoparticles is nonzero,
the collective electronic excitations couple to spin-dependent excitations in
the nanoparticle. \cite{yin09} This is the case in nanoparticles with an open
electronic shell, particularly for the case of ferromagnetic nanoparticles. 
It then becomes possible to affect the magnetization indirectly by exciting 
the charge degrees of freedom, and the influence of the magnetic field on the latter
might become important.

\begin{acknowledgments}
We thank Rodolfo Jalabert for fruitful discussions and for his careful
reading of the manuscript, and St\'ephane Berciaud, Fran{\c c}ois Gautier, and
Mircea Vomir for useful comments.
\end{acknowledgments}

\appendix
\section{Semiclassical calculation of the surface plasmon and surface magnetoplasmon linewidths}
\label{app:details}
In this Appendix, we present the details of our semiclassical calculation of the surface
(magneto)plasmon linewidths discussed in Sec.~\ref{sec:landau}. Introducing the notation
$\varepsilon_0=\hbar^2/2m_\mathrm{e}a^2$ and
$\kappa=\hbar\omega_\mathrm{c}/2\varepsilon_0$, we expand the semiclassical density of
states for fixed $l$ and $m$ [Eq.~\eqref{eq:DOS}] for
$\varepsilon\gg\hbar\omega_\mathrm{c}$ to obtain \cite{footnote:DOS}
\begin{equation}
\label{eq:DOS_approx}
\varrho_{l,m}(\varepsilon)\simeq\varrho_{l,0}(\varepsilon)\left[1+\kappa
mf_l(\varepsilon)\right],
\end{equation}
where 
\begin{equation}
\label{eq:f_l}
f_l(\varepsilon)=\frac{1}{\varepsilon/\varepsilon_0}-\frac{1}{2}\frac{1}{\varepsilon/\varepsilon_0-(l+1/2)^2}.
\end{equation}
With Eq.~\eqref{eq:DOS_approx} and using the expression of
the angular matrix elements \eqref{eq:angular}, the functions $F$ and $F_\pm$ [cf.\
Eq.~\eqref{eq:F}] that enter the surface plasmon and surface
magnetoplasmon linewidths [cf.\ Eq.~\eqref{eq:gammas}] read
\begin{subequations}
\label{eq:F_inter}
\begin{align}
F=&\;\frac{1}{3}
\int_{\max{(\varepsilon_\mathrm{F},\hbar\omega_\mathrm{M})}}^{\varepsilon_\mathrm{F}
+\hbar\omega_\mathrm{M}}\mathrm{d}\varepsilon\, \varepsilon(\varepsilon-\hbar\omega_\mathrm{M})
\sum_{l}\varrho_{l,0}(\varepsilon)
\nonumber\\
&\times\left[(l+1)\varrho_{l+1,0}(\varepsilon-\hbar\omega_\mathrm{M})+l\varrho_{l-1,0}(\varepsilon-\hbar\omega_\mathrm{M})\right],
\end{align}
and
\begin{align}
F_\pm=&\;\frac{1}{3}
\int_{\max{(\varepsilon_\mathrm{F},\hbar\omega_\pm)}}^{\varepsilon_\mathrm{F}
+\hbar\omega_\pm}\mathrm{d}\varepsilon\, \varepsilon(\varepsilon-\hbar\omega_\pm)
\sum_{l}\varrho_{l,0}(\varepsilon)\nonumber\\ 
&\times\Bigg\{
(l+1)\varrho_{l+1,0}(\varepsilon-\hbar\omega_\pm)
\Bigg[
1\mp\kappa
f_{l+1}(\varepsilon-\hbar\omega_\pm)
\nonumber\\
&\mp\frac{l}{2}\kappa\Big(f_l(\varepsilon-\hbar\omega_\pm)+f_{l+1}(\varepsilon-\hbar\omega_\pm)\Big)
\Bigg]
\nonumber\\
&+l\varrho_{l-1,0}(\varepsilon-\hbar\omega_\pm)
\Bigg[
1\mp\kappa
f_{l-1}(\varepsilon-\hbar\omega_\pm)
\nonumber\\
&\pm\frac{l+1}{2}\kappa\Big(f_l(\varepsilon-\hbar\omega_\pm)+f_{l-1}(\varepsilon-\hbar\omega_\pm)\Big)
\Bigg]
\Bigg\}.
\end{align}
\end{subequations}

Consistently with the semiclassical approximation (high-energy limit), we now
assume that $l\gg1$ (i.e., $l\simeq l\pm1$) and approximate in Eq.~\eqref{eq:F_inter} the summation over
$l$ by an integral. With Eqs.~\eqref{eq:DOS} and \eqref{eq:f_l}, we obtain
\begin{subequations}
\begin{align}
F=&\;\frac{1}{3(2\pi)^2}
\int_{\max{(\varepsilon_\mathrm{F},\hbar\omega_\mathrm{M})}}^{\varepsilon_\mathrm{F}
+\hbar\omega_\mathrm{M}}\mathrm{d}\varepsilon
\int_0^{\sqrt{(\varepsilon-\hbar\omega_\mathrm{M})/\varepsilon_0}}\mathrm{d}l\,
2l
\nonumber\\
&\times\sqrt{\left(\frac{\varepsilon}{\varepsilon_0}-l^2\right)
\left(\frac{\varepsilon-\hbar\omega_\mathrm{M}}{\varepsilon_0}-l^2\right)},
\end{align}
and
\begin{align}
F_\pm=&\;\frac{1}{3(2\pi)^2}
\int_{\max{(\varepsilon_\mathrm{F},\hbar\omega_\pm)}}^{\varepsilon_\mathrm{F}
+\hbar\omega_\pm}\mathrm{d}\varepsilon
\int_0^{\sqrt{(\varepsilon-\hbar\omega_\pm)/\varepsilon_0}}\mathrm{d}l\,
2l
\nonumber\\
&\times\sqrt{\left(\frac{\varepsilon}{\varepsilon_0}-l^2\right)
\left(\frac{\varepsilon-\hbar\omega_\pm}{\varepsilon_0}-l^2\right)}
\nonumber\\
&\times
\left\{
1\pm\kappa\left[\frac{1}{2}\frac{1}{(\varepsilon-\hbar\omega_\pm)/\varepsilon_0-l^2}
-\frac{\varepsilon_0}{\varepsilon-\hbar\omega_\pm}\right]
\right\}.
\end{align}
\end{subequations}
Expanding the above expressions for $\omega_\mathrm{c}\ll\omega_\mathrm{M}$, one
finds with Eq.~\eqref{eq:gammas} our final results for the surface
(magneto)plasmon linewidths, Eqs.~\eqref{eq:gamma} and \eqref{eq:gamma_pm}.

\section{Auxiliary functions for the surface plasmon and surface magnetoplasmon
linewidths}
\label{app:aux}
\begin{figure*}[tbh!]
\includegraphics[width=.46\linewidth]{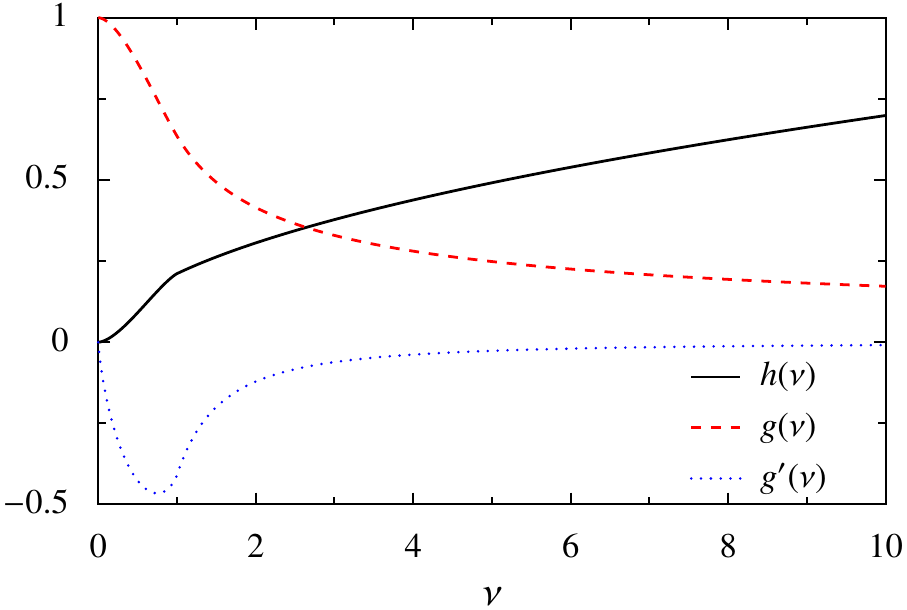}
\caption{\label{fig:auxiliary_functions}%
(Color online)
The function $g(\nu)$ of Eq.~\eqref{eq:g} (red dashed line), its derivative
$g'(\nu)$ (blue dotted line), and the function
$h(\nu)$ of Eq.~\eqref{eq:h} (black solid line).}
\end{figure*}

The function $g(\nu)$ entering the surface plasmon linewidth \eqref{eq:gamma} is
defined as
\begin{equation}
g(\nu)=\frac{2}{\nu}\int_{\max(1, \nu)}^{1+\nu}\mathrm{d}x
\int_0^{x-\nu}\mathrm{d}y \sqrt{(x-y)(x-y-\nu)}.
\end{equation}
The double integral can easily be evaluated, and one finds \cite{yanno92_AP}
\begin{subequations}
\label{eq:g}
\begin{align}
g(\nu)=&\;
\frac{1}{3\nu}\left[(1+\nu)^{3/2}-(1-\nu)^{3/2}\right]
\nonumber\\
&+\frac{\nu}{4}\left(\sqrt{1+\nu}-\sqrt{1-\nu}-\nu\ln{\nu}\right)
\nonumber\\
&+\frac{\nu}{2}\left[
\left(1+\frac{\nu}{2}\right)\ln{\left(\sqrt{1+\nu}-1\right)}
\right.\nonumber\\
&-\left.\left(1-\frac{\nu}{2}\right)\ln{\left(1-\sqrt{1-\nu}\right)}
\right]
\end{align}
for $\nu\leqslant1$ and
\begin{align}
g(\nu)=&\;
\frac{1}{3\nu}(1+\nu)^{3/2}
+\frac{\nu}{4}\left(\sqrt{1+\nu}-\ln{\nu}\right)
\nonumber\\
&+\frac{\nu}{2}\left[
\left(1+\frac{\nu}{2}\right)\ln{\left(\sqrt{1+\nu}-1\right)}
-\frac{\nu}{2}\ln{\sqrt{\nu}}
\right]
\end{align}
\end{subequations}
for $\nu>1$.
The function $g(\nu)$ is shown in Fig.~\ref{fig:auxiliary_functions} (red dashed line). 
Its asymptotic behaviors are 
$g(\nu)\simeq1+\nu^2[\ln{(\nu/4)}-1/6]/4$ and $g(\nu)\simeq8/15\sqrt{\nu}$ for
$\nu\ll1$ and $\nu\gg1$, respectively. 
Its derivative $g'(\nu)$
entering the surface magnetoplasmon linewidth \eqref{eq:gamma_pm} is shown
as a blue dotted line.

The function $h(\nu)$ entering the surface magnetoplasmon linewidths
\eqref{eq:gamma_pm} is defined as 
\begin{align}
h(\nu)=&\int_{\max(1, \nu)}^{1+\nu}\mathrm{d}x
\int_0^{x-\nu}\mathrm{d}y \sqrt{(x-y)(x-y-\nu)}
\nonumber\\
&\times\left[\frac{1}{2(x-y-\nu)}-\frac{1}{x-\nu}\right].
\end{align}
\begin{widetext}
Its explicit expression reads
\begin{subequations}
\label{eq:h}
\begin{align}
h(\nu)=&
-\frac{\nu}{2}
\Bigg\{
\sqrt{1+\nu}-\sqrt{1-\nu}
+\ln{\left(\frac{\sqrt{1+\nu}-1}{\sqrt{\nu}}\right)}
\left[1-\frac{\nu}{2}\ln{\left(\frac{\sqrt{1+\nu}-1}{\sqrt{\nu}}\right)}\right]
-\ln{\left(\frac{1-\sqrt{1-\nu}}{\sqrt{\nu}}\right)}
\left[1-\nu-\frac{\nu}{2}\ln{\left(\frac{1-\sqrt{1-\nu}}{\sqrt{\nu}}\right)}\right]
\nonumber\\
&+\nu\Bigg[\ln{\left(\frac{\sqrt{1+\nu}-1}{\sqrt{\nu}}\right)}\ln{\left(\frac{\sqrt{1+\nu}-1+\sqrt{\nu}}{\sqrt{\nu}}\right)}
-\ln{\left(\frac{1-\sqrt{1-\nu}}{\sqrt{\nu}}\right)}\ln{\left(\frac{1-\sqrt{1-\nu}+\sqrt{\nu}}{\sqrt{\nu}}\right)}
\nonumber\\
&+\mathrm{Li}_2\left(\frac{1-\sqrt{1+\nu}}{\sqrt{\nu}}\right)
-\mathrm{Li}_2\left(\frac{\sqrt{\nu}+1-\sqrt{1+\nu}}{\sqrt{\nu}}\right)
-\mathrm{Li}_2\left(\frac{\sqrt{1-\nu}-1}{\sqrt{\nu}}\right)
+\mathrm{Li}_2\left(\frac{\sqrt{\nu}+\sqrt{1-\nu}-1}{\sqrt{\nu}}\right)
\Bigg]
\Bigg\}
\end{align}
for $\nu\leqslant1$, and 
\begin{align}
h(\nu)=&
-\frac{\nu}{2}
\Bigg\{
\sqrt{1+\nu}
+\ln{\left(\frac{\sqrt{1+\nu}-1}{\sqrt{\nu}}\right)}
\left[1-\frac{\nu}{2}\ln{\left(\frac{\sqrt{1+\nu}-1}{\sqrt{\nu}}\right)}\right]
+\nu\Bigg[\ln{\left(\frac{\sqrt{1+\nu}-1}{\sqrt{\nu}}\right)}\ln{\left(\frac{\sqrt{1+\nu}-1+\sqrt{\nu}}{\sqrt{\nu}}\right)}
\nonumber\\
&+\mathrm{Li}_2\left(\frac{1-\sqrt{1+\nu}}{\sqrt{\nu}}\right)
-\mathrm{Li}_2\left(\frac{\sqrt{\nu}+1-\sqrt{1+\nu}}{\sqrt{\nu}}\right)
+\frac{\pi^2}{12}
\Bigg]
\Bigg\}
\end{align}
\end{subequations}
\end{widetext}
for $\nu>1$.
It involves the dilogarithmic function
\begin{equation}
\mathrm{Li}_2(z)=\sum_{k=1}^\infty\frac{z^k}{k^2}=\int_z^0\mathrm{d}t\,\frac{\ln{(1-t)}}{t}.
\end{equation}
The asymptotic behaviors of the function $h(\nu)$ for $\nu\ll1$ and $\nu\gg1$
read $h(\nu)\simeq-\nu^2[\ln{(\nu/4)}+1]/4$ and $h(\nu)\simeq2\sqrt{\nu}/9$,
respectively. The function $h(\nu)$ is plotted in
Fig.~\ref{fig:auxiliary_functions} (black solid line).



\begin{thebibliography}{99}
\bibitem{footnote:cgs}
Throughout the paper, we use c.g.s.\ units.
\bibitem{footnote:N}
We assume here that the number of valence electrons is large
($N\gg1$), such that $N-1\simeq N$. Therefore, the number of relative degrees of
freedom equals the total one.
\bibitem{footnote:confinment}
This approximation is well justified for nanometer-sized particles: The
evanescent part of a wave function of energy $\varepsilon<V_0$ can be neglected
when $\sqrt{2m(V_0-\varepsilon)}a/\hbar \gg 1$. In typical nanoparticles one
has $V_0-\varepsilon\simeq\varepsilon_\mathrm{F}$ 
such that $V_0\to\infty$ becomes
allowed when $k_\mathrm{F}a\gg 1$, with $k_\mathrm{F}$ being the Fermi wave vector. 
This is fulfilled since
$k_\mathrm{F}\simeq\unit[10]{nm^{-1}}$ for metals.
\bibitem{footnote:temperature}
The electronic temperature $T$ of the system is much smaller than the Fermi
temperature $T_\mathrm{F}$ which is of the order of $10^4$ to $\unit[10^5]{K}$
for typical metals. Thus, the effect of temperature on the surface
(magneto)plasmon linewidths can be neglected here. For its effect in the case of the
usual surface plasmon linewidth, where finite-temperature corrections are of
order $(T/T_\mathrm{F})^2$, see Ref.~\onlinecite{weick06_PRB}. The temperature
correction for the surface magnetoplasmon linewidths [see
Eq.~\eqref{eq:gamma_pm}] is expected to be the same.
\bibitem{footnote:oscillations}
Higher-order corrections in $\hbar$ to the density of states 
lead to an oscillation of the surface plasmon lifetime as a function of the 
size of the nanoparticle (Refs.~\onlinecite{molin02_PRB}, \onlinecite{weick05_PRB}). 
This effect is only relevant for very small nanoparticles 
($a\lesssim\unit[1.5]{nm}$), and we disregard it here.
\bibitem{footnote:observability}
The $1/a$ size dependence of the decay rates \eqref{eq:gamma_pm}
suggests that taking larger nanoparticles would facilitate the observation of
two well-defined surface magnetoplasmon lines in the absorption spectrum.
However, for large nanoparticles ($a\gtrsim\unit[10]{nm}$), the radiation
damping decay rate $\gamma_\mathrm{rad}\propto a^3$ increases above the Landau
damping (Ref.~\onlinecite{kreibig}) and masks the two collective resonances.
\bibitem{footnote:DOS}
The total density of states
$\varrho(\varepsilon)=\sum_{lm}\varrho_{l,m}(\varepsilon)$ corresponds to the leading
order Weyl term of the classical density of states
(Ref.~\onlinecite{weick05_PRB}) and is thus independent of
the magnetic field (Refs.~\onlinecite{richt96_PR}, \onlinecite{tanak96_PRB}).
\end{thebibliography}
\end{document}